\documentclass[12pt]{article}

\usepackage[english]{babel}
\usepackage[T1]{fontenc}
\usepackage[utf8]{inputenc}
\usepackage{amssymb,amsmath,array,hhline, dsfont}
\usepackage{latexsym}
\usepackage{graphicx}
\usepackage{verbatim}
\usepackage{stmaryrd}

\title{On and off-diagonal Sturmian operator: dynamic and spectral dimension}

\author{Laurent Marin
\\
\\
{\small Department Mathematik, Universit\"at Erlangen-N\"urnberg, Germany}
}

\date{ }

\newtheorem{theo}{Theorem}
\newtheorem{defini}{Definition}
\newtheorem{proposi}{Proposition}
\newtheorem{lemma}{Lemma}
\newtheorem{coro}{Corollary}
\newtheorem{rem}{Remark}

\newcommand{\CM}{{\mathbb C}}

\newcommand{\RM}{{\mathbb R}}

\newcommand{\ZM}{{\mathbb Z}}

\newcommand{\e}{\varepsilon}

\begin{document}

\maketitle

\begin{abstract}
We study two versions of quasicrystal model, both subcases of Jacobi matrices. For off-diagonal model, we show an upper bound of dynamical exponent and the norm of the transfer matrix. We apply this result to the off-diagonal Fibonacci Hamiltonian and obtain a sub-ballistic bound for coupling large enough. In diagonal case, we improve previous lower bounds on the fractal box-counting dimension of the spectrum.  
\end{abstract}

\vspace{.5cm}

\section{Sturmian on and off-diagonal models }


On and off-diagonal models are special cases of the Jacobi operators.  
Given two real-valued sequences $a$ and  $b$, a Jacobi operator $H$ acts on $\ell^2 (\ZM)$ in the following way
\begin{equation}\label{jacobidef}
H \psi(n) = a(n) \psi(n+1) + a(n-1) \psi(n-1) + b(n) \psi(n), \quad n \in \ZM.
\end{equation}
$H$ is associated to a self-adjoint tridiagonal matrix with diagonal entries filled with values of $b$ and off-diagonal entries filled with values of $a$. 

Let $\beta \in [0,1]$ be an irrational number, on and off-diagonal Sturmian models will be defined respectively by setting in \eqref{jacobidef}
\begin{equation}\label{sturmon}
a(n) = 1, \quad  b(n) = \lambda_1\Bigl( \lfloor (n+1) \beta \rfloor - \lfloor n \beta \rfloor \Bigr) 
\end{equation}
respectively
\begin{equation}\label{sturmoff}
a(n) = (\lambda_1-\lambda_2)\Bigl(\lfloor (n+1)\beta \rfloor - \lfloor n\beta \rfloor \Bigr) + \lambda_2, \quad b(n) = 0, \quad n \in \ZM
\end{equation}

where $\lambda_1, \,\lambda_2 \in \mathbb R^+$ and for $x \in \mathbb R$, $\lfloor x \rfloor$ denotes the largest integer smaller than $|x|$. The term Sturmian refers to quasiperiodicity of sequences $b$ in \eqref{sturmon} and $a$ in \eqref{sturmoff}. On-diagonal model is more usually named discrete Schr\"{o}dinger operator and when $\beta$ is the golden mean $\frac{\sqrt{5}-1}{2}$, the Sturmian model is oftenly called the Fibonacci Hamiltonian.


The study of those models was initiated with the introduction  of the Fibonacci operator in the early
1980's by Kohmoto et al.  \cite{kohmoto} and Ostlund et al. \cite{ostlund}. At that point in time, the
interest in this model was based mainly on the existence of an exact renormalization
group procedure, the appearance of critical eigenstates and zero measure Cantor
spectrum. Only shortly thereafter, Shechtman et al. \cite{she}  reported their discovery of
structures, now called quasicrystals, whose are shown to have aperiodic structure.
Sturmian sequences are central models of a quasicrystal in one dimension. Indeed, it is aperiodic  moreover, it belongs to
virtually all classes of mathematical models of quasicrystals that have since been
proposed. We refer to the reader to \cite{baake} for a recent account of the mathematics related to the modelling and study of quasicrystals.
Thus, the study of Sturmian operators was further motivated by the interest in electronic spectra and transport properties of one-dimensional quasicrystals.
 It is shown that those operators give rise to anomalous transport for a large class of irrational number \cite{dam1, LaurentThesis, Laurentarticle}. Moreover, it  exhibits a number of interesting phenomena, such as Cantor spectrum of Lebesgue measure zero \cite{s1,s3,BIST1989} and purely singular continuous spectral measure \cite{DamLenz, DKL, BIST1989}.
Consequently, apart from the almost Mathieu operator, the Fibonacci operator has
been the most heavily studied quasi-periodic operator in the last three decades;
compare the survey articles \cite{dam5, dam6, s2}.
Partly due to the choice of the model in the foundational papers \cite{kohmoto,ostlund} the
mathematical literature on the Fibonacci operator has so far only considered the
diagonal model. Given the connection to quasicrystals
and hence aperiodic point sets, and particularly cut-and-project sets, it is however
equally (if not more) natural to study  the off-diagonal Sturmian model.  We do
refer the reader to \cite{baake} for background. For further motivation to study also the off-diagonal model,
we mention that it has been the object of interest in a number of physics  and then mathematics recent papers \cite{even, EL2, Janine, damgorod, KST, VP}.

In this paper, we are interested in transport properties and in fractal dimension of the spectrum. More precisely, the reader will find in the following:

For off-diagonal Sturmian model,
\begin{itemize}

\item We show pseudo spectrum and spectrum are equal (Theorem \ref{main3}).

\item We established a link between outside probabilities and transfer matrices in both time-averaged (Theorem \ref{main}) and non time-averaged settings (Theorem \ref{main2}).  

\item In the Fibonacci case, we deduce from previous item a dynamical upper bound for the wavepacket spreading (Theorem \ref{thprincipal}). This bound is  sub-ballistic for hopping constants $\lambda_1,\lambda_2$ well choosen.  

\end{itemize}

For on-diagonal Sturmian model,

\begin{itemize}
\item We improve previous lower bound for the box-counting dimension of the spectrum (\cite{Laurentarticle}).  This bound is valid for a large class of irrational number verifying a Lebesgue measure $1$ diophantine condition (Theorem \ref{thbornesur}).

\item Considering a smaller class of irrational number, but still Lebesgue measure 1, we are able to improve our previous lower bound (Theorem \ref{bornepresquesure}).   

\end{itemize}

\section{The trace map application}


Denote the continued fraction expansion of $\beta$ by
$$
\beta = [a_1, a_2, \dots]= \frac{1}{a_1+\frac{1}{a_2 + \dots}}
$$
and define the  best rational  approximants,
$$
p_{k+1} = a_{k+1} p_k + p_{k-1},\qquad p_0 = 1,\qquad  p_{-1}=0,
$$
$$
q_{k+1} = a_{k+1} q_k + q_{k-1},\qquad q_0 = 0,\qquad q_{-1}=1.
$$


A classic tool to investigate one-dimensional model is to write to free equation
\begin{equation}\label{freeeq}
H\psi (n) =  E \psi(n)
\end{equation}
where $E$ is a real or a complex number.
Equation \eqref{freeeq} can be rewritten in both on and off-diagonal case respectively as 
$$
\Psi^b (n) = \binom{\psi (n+1)}{\psi(n)} =  \begin{pmatrix}
           E - b(n) & -1 \\
	   1 & 0
	 \end{pmatrix}
 \binom{\psi(n)}{\psi(n-1)}
$$
$$
\Psi^a (n) = \begin{pmatrix} \psi (n+1) \\ a(n)\psi(n) \end{pmatrix} = \frac{1}{a(n)} \begin{pmatrix} E & -1 \\ a(n)^2 & 0 \end{pmatrix} \begin{pmatrix} \psi (n) \\ a(n-1)\psi(n-1) \end{pmatrix}
$$
Denote $T^b_n(E) =  \begin{pmatrix}
           E - b(n) & -1 \\
	   1 & 0
	 \end{pmatrix}$ and $T_n^a (E) = \frac{1}{a(n)} \begin{pmatrix} E & -1 \\ a(n)^2 & 0 \end{pmatrix} $. In all the following, objects with exponent $a$ are associated to the off-diagonal model, while those with exponent $b$ are associated to diagonal case. When there is no ambiguity, we will drop it.
Then for $i = a,b$ and with $F_n^i(E) = T_n^i(E)\dots T_1^i(E)$, one constructs recursively a solution $\psi^i$ of the equation (\ref{freeeq}) with
$$
\Psi^i (n) = F_n^i(E) \Psi^i(0).
$$


We define the so-called transfer matrices, denoted by $M_n^i(E) = F_{q_n}^i(E)$.
One important objet is the traces of tranfer matrices, denote $x_n^i = \mathrm{Tr} M_n^i$ and $z_n^i = \mathrm{Tr} (M_{n-1}^iM_n^i)$. In the following, we will drop the index $i$ whenever it is possible for sake of simplicity.

The transfer matrix $k-$evolution follows the simple rule (see e.g. \cite{BIST1989,ioc1,r1})
\begin{equation}\label{matriceevol}
M_k = M_{k-2} M_{k-1}^{a_k},\quad k \geq 1.
\end{equation}
We can extend this relation to $k=0$ by setting $ M_{-1}^a = \begin{pmatrix} \frac{\lambda_2}{\lambda_1} & 0 \\ 0 &\frac{\lambda_1}{\lambda_2} \end{pmatrix}$ and $M^b_{-1} = \begin{pmatrix} 1 & -\lambda_1 \\ 0 & 1 \end{pmatrix}$.

The evolution of these sequences is derived from taking the trace in (\ref{matriceevol})  and by Cayley-Hamilton Theorem (see \cite{BIST1989} for details); for $k\geq 0$
\begin{equation}
\label{traceevolx}
x_{k+1} = z_k S_{a_{k+1}-1}(x_k) - x_{k-1} S_{a_{k+1}-2}(x_k),
\end{equation}
\begin{equation}\label{traceevolz}
z_{k+1} = z_k S_{a_{k+1}}(x_k) - x_{k-1} S_{a_{k+1}-1}(x_k), 
\end{equation}
where the $S_l$ are Chebyshev second order polynomials
$$
S_{l+1}(x) = x S_l(x) - S_{l-1}(x), \quad \forall l \geq 0, \qquad S_{0}(x) =1,
\qquad S_{-1}(x) =0.
$$

The initial conditions of these two sequences are given by, (see \cite{damgorod, Janine, BIST1989})
\begin{equation}
\label{condinit}
x_{-1}^a= \frac{\lambda_1^2+\lambda_2^2}{\lambda_1\lambda_2}, \quad x^a_0 = \frac{E}{\lambda_2}, \quad  z^a_0 = \frac{E}{\lambda_1},
\end{equation}
\begin{equation}
\label{condinit2}
x_{-1}^b= 2,\qquad  x^b_0 = E,\qquad  z^b_0 = E - \lambda_1.
\end{equation}

Sequences $x^i$ and $z^i$ verify the Fricke-Vogt invariant, namely for $k \geq 0$ and $E$, one has
\begin{equation}
\label{invariant}
(x_k^i)^2 + (x_{k+1}^i)^2 + (z_{k+1}^i)^2 -x_k^i x_{k+1}^i z_{k+1}^i = (c^i)^2  +4
\end{equation}
with $c^a = \frac{(\lambda_1-\lambda_2)(\lambda_1+\lambda_2)}{\lambda_1\lambda_2}
$
 and  $c^b = \lambda_1$.

\section{Periodic approximants for off-diagonal Sturmian model}

We define the following sequence of sets
$$
\sigma_{n,k} = \{ E \in \RM \;{\bf : }\; |\mathrm{Tr}  M_{n-1} M_{n}^k| \leq 2 \}.
$$
In case $k=0$, those sets are called the periodic approximants of $H$ as $\sigma_{n,0}$ is the spectrum  of a periodic operator defined as in \eqref{jacobidef} with a $q_n$-periodic sequence $a$, see \cite{s1,BIST1989}.

The pseudo spectrum is defined as  $B_\infty = \{ E \in \RM \;{\bf : }\; \{x_n (E)\}_n \text{ is bounded } \}$
and  denote $N_0(E)$ the first index such that $|x_{N_0}(E)|\leq 2$.

\begin{lemma}
\label{explosion1}

If $|x_0| >2$ and $|z_0|>2$, then the two sequences $\{x_n,z_n\}_n$ are unbounded. 

\end{lemma}

\noindent {\bf Proof.} Let us first check the following induction, suppose that at step $n$, we have $|z_n|>2$, $|x_n|>2$ and $|z_n|>|x_{n-1}|$. Then,  (\ref{traceevolz}) implies that for some integer $l>0$,
\begin{align}
\notag |z_{n+1}| &= |z_n S_l (x_n) - x_{n-1} S_{l-1}(x_n)| \\ \notag 
         &\geq |z_n| |S_l(x_n)-S_{l-1}(x_n)| \\ 
        &\geq |z_n| (|x_n|-1)^l.  \label{20}
\end{align}
Since $|x_n|,|z_n|>2$, one has   $|z_{n+1}|>|z_{n}|$ and  $|z_{n+1}|>|x_{n}|$.
From (\ref{traceevolx}), the same argument gives   $|x_{n+1}|>|z_{n}|$ completing the induction.
Clearly if the induction is fulfilled, it implies that both sequences are increasing and taking logarithme in (\ref{20}) show they are unbounded.

We check now the initial condition of the induction.
Let first assume that $\lambda_1<\lambda_2$. 
Then $|x_0=\tfrac{E}{\lambda_2}|>2$ implies that $|z_0=\tfrac{E}{\lambda_1}|>\frac{2\lambda_2}{\lambda_1} > \frac{\lambda_2}{\lambda_1} + \frac{\lambda_1}{\lambda_2}=x_{-1}$. 
Now suppose that $\lambda_2<\lambda_1$, we argue that $|z_1|>|x_0|$ and $|z_1|, |x_1|>2$.
First if $a_1=1$, then 
$$
z_1 = z_0 x_0 -x_{-1}
$$
and since $|x_0|>|x_{-1}|$ and $|z_0|>2$, we have that $|z_1|>|x_0|$. Moreover, in that case $x_1=z_0$ completing the proof.
Then suppose that $a_1=l>1$. The assumption $\lambda_2<\lambda_1$ with $|x_0|,|z_0|>2$ implies that $|E|>2\lambda_1$. Then there exists a positive constant $\e$ such that $E=\pm(2\lambda_1+\e)$,
\begin{align*}
|z_1| &= |(z_0x_0-x_{-1}) S_{l-1}(x_0) - z_0 S_{l-2}(x_0)| \\
&= \left|\left(\frac{E^2}{\lambda_1\lambda_2}-\frac{\lambda_1^2+\lambda_2^2}{\lambda_1\lambda_2}\right) S_{l-1}(x_0) - \frac{E}{\lambda_1} S_{l-2}(x_0)\right| \\
&= \left|\left(\frac{3\lambda_1}{\lambda_2} -\frac{\lambda_2}{\lambda_1}+\frac{\e^2}{\lambda_1\lambda_2}+ \frac{4\e}{\lambda_2}\right) S_{l-1}(x_0) \mp \left(2+ \frac{\e}{\lambda_1}\right) S_{l-2}(x_0)\right| \\
&\geq \left(\frac{2\lambda_1}{\lambda_2}+\frac{\e^2}{\lambda_1\lambda_2}+ \frac{4\e}{\lambda_2}\right)|S_{l-1}(x_0)-S_{l-2}(x_0)| \\
&\geq \left(\frac{2\lambda_1}{\lambda_2} +c_{\e}\right) (|x_0|-1)^{l-1} \\
&\geq 2|x_0|-2 > |x_0|.
\end{align*}
The same argument upon $x_1$ implies $|x_1|>2\frac{\lambda_1}{\lambda_2}+c_{\e}>2$ completing the proof.
\hfill $\Box$

\begin{lemma}\label{lemma5}
Let  $x_n$ and $z_n$ be the trace map sequence define in \eqref{traceevolx} and \eqref{traceevolz}, $\delta >0$. 

If there exists  an integer $N$ such that
\begin{equation}\label{conditionexplosion}
|x_{N-1}| \leq 2+\delta   ,\quad |x_N| > 2+\delta  \hspace{3mm} and
\hspace{3mm} |z_{N}| > 2+\delta,
\end{equation}
then modulus of the two sequences are superexponentially increasing
$$
|x_{k+1}| \geq |z_k| \geq e^{c(\delta) G_{k-N}} +1 \hspace{5mm},  \forall k > N,
$$
with
$$
G_k = G_{k-1} + a_k G_{k-2},\quad G_0=1,\quad G_{-1}=1.
$$

Moreover, 
\begin{equation}
\label{Binfini}
B_\infty = \bigcap_{n=N}^{\infty} (\sigma_{n,0} \cup \sigma_{n+1,0}), \text{ any } N \geq N_0.
\end{equation}
\end{lemma}

\noindent {\bf Proof.} First part of the Lemma follows from \eqref{traceevolx}-\eqref{traceevolz} (see \cite{Laurentarticle, LaurentThesis} for a detailed Proof). The occurence of $\delta$ plays no role in the result and will be needed for technical reason in section 5. The condition upon $z_N$ could also be replaced by $|x_{N+1}|>2$, see \cite{BIST1989} for this proof version. 

For the second part, it remains to show is that for $E \in B_\infty$, $N_0 (E)$ is finite, then the result follows from the first part of the lemma and identical argument that in the on-diagonal case (see \cite{s1, BIST1989}).
Suppose that $x_n$ is bounded and greater than two in modulus.
This impling at any rank $n$, that $2<|x_{n+1}|\leq |x_{n-1}|$. Elsewhere, replacing the hypothesis $|x_{n-1}|\leq 2<|x_{n+1}|$ by the weaker one $|x_{n-1}|<|x_{n+1}|$ in Proposition 4 in \cite{BIST1989}, one can show that the sequence $x_n$ is growing exponentially.
Thus each subsequence on even and odd index has to be decreasing and bounded from below by 2, and thus has a limit $l_1$. Then the Fricke-Vogt invariant implies the sequence $z_n$ also has a limit denoted $l_2$. Therefore the point $(l_2,l_1,l_1)$ is a fixed point of the trace map and the  set of equation coming from $T(l_2,l_1,l_1) = (l_2,l_1,l_1)$ should be verified. This is easily shown to be impossible using the fact that $|S_a (l_1)| \geq 2$ since $|l_1| \geq 2$.  
\hfill $\Box$

\begin{theo}\label{main3}
For any $\lambda_1, \lambda_2>0$,  $\sigma(H)$ the spectrum of $H$ and  $ B_\infty$ coincide.
\end{theo}

\noindent {\bf Proof.} Using Lemma \ref{lemma5}, one can argue the same as in \cite{s1}, replacing $N\geq 0$ by $N\geq N_0$ anytime it appears. 
\hfill $\Box$

\section{Dynamical upper bound for off-diagonal model}


In this section, we are mostly interested in dynamical properties of the wavepacket.
It is known for this type of aperiodic model that the unitary group
$
\psi (t) = e^{-itH}\psi (0)
$
will spread with time $t$. Here, $\psi(0)$ is some well localized initial condition, for example, $\delta_1$ the Dirac function on the first site of $\ell^2 (\ZM )$.
We want to quantify this spreading, thus we recall  one usual way to measure it.

Denote by $a(n,t) = |\langle e^{-itH}\delta_1,
\delta_n \rangle |^2$ the probability for the system to be at the $n^{th}$ site at time $t$. Here  $\{ \delta_i\}_{i \in \ZM}$ is the canonical basis of $\ell^2(\ZM)$.

We denote the outside probabilities by
$$
P(N,t) = \sum_{|n| > N} a(n,t),
$$
and 
$$
P_r(N,t) = \sum_{n > N} a(n,t), \hspace{5mm} P_l(N,t) = \sum_{n < -N} a(n,t)
$$
Namely, $P(N,t)$ is the probability to be outside the ball of size $N$ at time $t$.

For all
$\alpha \in [0,+\infty ] $, as it is done in \cite{ger1}, define
$$
S^-(\alpha) = -\liminf_{t \to \infty } \frac{\ln P(t^\alpha -2 ,t)
}{\ln t}, \quad
S^+(\alpha) = -\limsup_{t \to \infty } \frac{\ln P(t^\alpha -2 ,t)
}{\ln t}.
$$

The following critical exponents are particular of interest:
$$
\alpha^{\pm}_l = \sup \{\alpha \geq 0: S^\pm (\alpha) =0\},
\quad
\alpha^{\pm}_u = \sup \{\alpha \geq 0: S^\pm (\alpha) < \infty \}.
$$

They verify $0\leq \alpha_l^\pm \leq \alpha_u^\pm$. In particular, if $\gamma
> \alpha_u^+$ then $P(t^\gamma,t)$ goes to $0$ faster than polynomially.
$\alpha^{\pm}_l$ can be interpreted as the (lower and upper) rates of
propagation of the essential part of $\psi$ and $\alpha^{\pm}_u$
as  the rates of propagation of the fastest part of $\psi$ (see \cite{ger1} ). 
 
It is also convenient sometimes to consider these definitions in average in time. 
We define the time-averaged probability
$\langle a(n,T) \rangle = \frac{2}{T} \int_0^\infty e^{-2t/T} a(n,t) dt$.
Replacing then $a(n,t)$ by $\langle a(n,T) \rangle$ we can then define its time-averaged  outside probabilities $\langle P(N,T) \rangle$ and all the exponents above, denoted with a tilda.

Now main notations are set up, we extent the link between (time-averaged or not) outside probabilities and transfer matrices to off-diagonal models and show the following:

\begin{theo}\label{main}
Suppose $H$  defined in \eqref{jacobidef} with $a,b$ defined in \eqref{sturmoff}, and let  $K \geq 4$ be such that $\sigma(H) \subset [-K+1, K-1]$. Then the time-averaged outside probabilities can be bounded from above in terms of transfer matrix norms as follows:
$$
\langle P(N,T) \rangle \lesssim exp(-cN) +T^3 \int_{-K}^{K} \left( \max_{1\leq n \leq N} \| F (n,E + \tfrac{i}{T}) \|^2 \right)^{-1}dE.
$$
\end{theo}

 The proof starting point, as in diagonal case (see   \cite{dam1} ), is Parseval formula,
$$
\langle a(n,T) \rangle \; = \; \frac{1}{T\pi} \int_{-\infty}^{\infty} |\langle (H-E-\tfrac{i}{T})^{-1} \delta_1, \delta_n \rangle |^2 dE.
$$
Denote $\varepsilon = \frac{1}{T} $ and $R(z) = (H-zI)^{-1}$ for $z \in \mathbb{C}\backslash \mathbb{R}$. Let us assume that $T > 1$ and therefore $ 0< \varepsilon <1$.
Then, we have

\begin{equation}\label{g4}
\langle P_r (N,T) \rangle \; = \;  \frac{\varepsilon}{\pi} \int_{-\infty}^{\infty} M_r(N,E+i \varepsilon)dE,
\end{equation}
where
$$
M_r(N,z) \; = \; \sum_{n>N} | \langle R(z) \delta_1, \delta_n \rangle |^2 \; = \; \| \chi_N R(z)\delta_1 \|^2,
$$
and $\chi_N(n) =0, \, n \leq N,  \; \chi_N(n)=1, \, n \geq N+1$. We need to bound $M_r(N,z)$ from above. If the energy $E$ is outside the spectrum of $H$, one can use Combes-Thomas estimate (see e.g. \cite{combes}). Assume that $\eta =$dist$ ( E+i\varepsilon, \sigma (H)) \geq 1$, then we have,
$$
| \langle R(z)\delta_1 , \delta_n \rangle | \; \leq \; \frac{2}{\eta} \exp (-d \min \{
d\eta,1 \}|n-1|)
$$
with some universal positive constant $d$.
Using this estimate, one can see easily that 
$$
 \int_{E:|E|\geq K} M_r (N,E+i\varepsilon)dE \; \leq \; C(K)\exp(-dN), \;d>0,
$$
and thus the limit goes to $0$ for any $N(T)=T^\alpha, \;\alpha>0.$ The remaining problem is to estimate the other part of the integral where $E+i\varepsilon$ may be $\varepsilon -$close to the spectrum of $H$
$$
L_r (N,\varepsilon) = \int_{-K}^K M_r (N,E+i\varepsilon)dE.
$$

We link $M_r(N,z)$ to the complex solutions to the stationary equation $H\psi=z\psi$.

First consider the truncated operator $H_N$, namely
\begin{equation}
H_N \psi (n) = w(n+1) \psi (n+1) + w(n) \psi(n-1)
\end{equation}
where 
$$
w(n) = a(n), \;  n \leq N, \quad w(n) = \lambda_2, \; n>N.
$$
Denote by $R_N(z)$ the resolvent of the operator $H_N$ and define 
$$
S(N,z) = \| \chi_N R_N (z) \delta_1 \|^2.
$$

\begin{lemma}\label{lem1}
For any $E \in [-K,K]$, $0< \varepsilon <1$, we have 
$$
M_r (N,E+i\varepsilon) \lesssim \varepsilon^{-2} S(N,E+i\varepsilon),
$$
where the implicit constant depends only on $\lambda_1,\lambda_2$.
\end{lemma}

\noindent {\bf Proof.}
Using the revolvent identity,
$$
R(z)\delta_1 - R_N(z)\delta_1 = R(z)\chi_N (H-\lambda_2\Delta) R_N(z) \delta_1= R(z)
\chi_N (d T^{\pm 1}) R_N(z) \delta_1 .
$$
where $d$ is a constant, namely $\lambda_1-\lambda_2$ or $0$ and $T$ is the shift operator , $T\psi(n)=\psi(n+1)$ acting on $\ell^2(\mathbb{Z})$.

Since $\| R(z) \| \leq \varepsilon^{-1}, \varepsilon < 1,$ and the sequence $a(n)$ is  bounded, we get
\begin{align*}
M_r(N,z) &\leq  2 S(N,z) + 2\| R(z)(\chi_N H-\lambda_2\Delta) R_N(z) \delta_1 \|^2 \\
         &\leq  2 S(N,z) + 2\varepsilon^{-2}\|\chi_N (H-\lambda_2\Delta) R_N(z) \delta_1 \|^2 \\
  &\leq  2 S(N,z) + 2\varepsilon^{-2}\|\chi_N (d T^{\pm 1}) R_N(z) \delta_1 \|^2 \\
&\leq C(d)\varepsilon^{-2} S(N,z). \hspace{8cm} \Box
\end{align*}

The next step is to link the quantity $S(N,z)$ with the solutions to the stationary equation. For any complex $z$, define $u_0 (n,z) $ as a solution to $Hu_0= zu_0, \; u_0(0,z)=0, \; u_0(1,z)=1$, and define  $u_1 (n,z) $ as a solution to $Hu_1= zu_1, \; u_1(0,z)=1, \;  u_1(1,z)=0$. Since $H$ and $H_N$  coincide for all $n \leq N$, their solutions $u_0$, $u_1$ coincide for all $n \leq N+1$. We will consider $u_0,u_1$ only for such $n$, thus we will use the same notation for $u_0,u_1$ for $H$ and $H_N$.

Now consider the free equation upon $H_N$
\begin{equation}
\label{34}
\lambda_2 u(n-1) + \lambda_2 u(n+1) \; = \; z u(n), \qquad n>N
\end{equation}   
and define 
$$
s_{1,2} \; = \; \frac{z\pm \sqrt{z^2-4\lambda_2^2}}{2\lambda_2}
$$
the roots of the caracteristic polynom of the sequence $u(n)$. Since $s_1 s_2 =1 $, one has $|s_1|>1$ and $|s_2|<1$.
Therefore $s_{1},s_2$ are the eigenvalues of equation (\ref{34}) whose general solution is
$$
u(n) = c_1 s_1^n + c_2 s_2^n.
$$

\begin{lemma}\label{lem2}
Let $z=E+i\e$, where $E \in [-K,K]$, $0 < \e < 1$. Then, for $n \ge N \ge 3$, we have
\begin{equation}\label{g6a}
|\langle R_N (z) \delta_1 , \delta_n \rangle| \le 2 \e^{-1} \frac{ |s_2
(z)|^{n-N}}{ |s_2 (z) u_0(N,z) -u_0(N+1,z)| },
\end{equation}
and
\begin{equation}\label{g6b}
|\langle R_N (z) \delta_1 , \delta_n \rangle| \le \e^{-1} \frac{ |s_2
(z)|^{n-N}}{ |s_2 (z) u_1(N,z) -u_1(N+1,z)| }.
\end{equation}
\end{lemma}

\noindent {\bf Proof.} The following formula holds for any off-diagonal model, 
(see \cite{KKL}):
\begin{equation}\label{g31a} \langle R(z) \delta_1, \delta_n \rangle=d(z) u_0
(n,z) +b(z) u_1 (n,z), \ n \ge 1,
\end{equation}
\begin{equation}\label{g31b}
\langle R(z) \delta_1, \delta_n \rangle=d(z) u_0 (n,z) +c(z) u_1
(n,z), \ n < 1,
\end{equation}
where
\begin{equation}
b(z)  = \frac{m_-(z)}{b(m_+(z)+m_-(z))}, \qquad
c(z)  = \frac{-m_+(z)}{b(m_+(z)+m_-(z))}, \qquad
d(z)  = \frac{-m_+(z)m_-(z)}{b(m_+(z)+m_-(z))}, 
\end{equation}
with some complex functions $m_+(z), m_-(z)$, called the $m$-functions which depend on the sequences $a$ and $b$. Since
$\|R(z) \delta_1\| \le \e^{-1}$ and
$$
d(z)=\langle R(z) \delta_1, \delta_1 \rangle , \qquad \  c(z)=\langle R(z) \delta_1, \delta_0
\rangle,
$$
we get $|d(z)| \le \e^{-1}, \; \ |c(z)| \le \e^{-1}$. Since $b(z)=1+c(z)$ and $\e \le 1$, we
also get $|b(z)| \le 2 \e^{-1}$. Summarizing,
\begin{equation}\label{g102}
|d(z)| \le
\e^{-1}, \qquad |c(z)| \le \e^{-1}, \qquad |b(z)| \le 2 \e^{-1}.
\end{equation}
One should stress that the bounds \eqref{g102} hold for any operator and the constants are universal.

Consider the operators $H_N$ and define $\phi=R_N (z)\delta_1$ (the vector
$\phi$ depends on $N$, of course). Since
$$
(H_N-z) \phi=\delta_1,
$$
the function $\phi (n)=\langle \phi, \delta_n \rangle$ obeys the equation
$H_N \phi(n)=z \phi(n),\quad n \ge 2$.  Since $w(n)= \lambda_1,  \ n\ge N+1$, $\phi(n)$ obeys
the free equation \eqref{34} for $n \ge N+1$. Hence,
\begin{equation}\label{g8}
(\phi
(N+k+1), \phi (N+k))^T \;= \; c_1 s_1 (z)^k e_1+c_2 s_2 (z)^k
e_2, \quad k \ge 0.
\end{equation}
Here $e_{1,2}=(s_{1,2} (z), 1)^T$ are two eigenvectors of the matrix
corresponding to the equation \eqref{34}, and the constants $c_1,c_2$ are
defined by
$$
(\phi (N+1), \phi(N))^T=c_1 e_1+c_2 e_2.
$$
Since $|s_2 (z)|<1, \ |s_1 (z)|>1$ and $\phi  \in \ell^2(\ZM)$,
the identity \eqref{g8} implies that $c_1=0$, and thus
\begin{equation}\label{g9}
(\phi (N+1), \phi (N))^T=D_N(z) (s_2 (z), 1)^T, \ D_N(z) \ne 0.
\end{equation}
On the other hand, \eqref{g31a} implies
\begin{equation}\label{g10a}
\phi^\pm
(N+1)\; = \;d_N (z) u_0(N+1, z) + b_N (z) u_1(N+1, z),
\end{equation}
\begin{equation}\label{g10b}
\phi (N)\; =\; d_N (z) u_0(N, z) + b_N (z) u_1(N, z).
\end{equation}
Remark that $d_N (z) \ne 0$, since it is a Borel transform of the spectral measure
corresponding to the vector $\delta_1$ \cite{KKL}. The same is true for $b_N (z),
c_N (z)$ since $m_+(z), m_-(z)$ are functions with  positive imaginary part (it also
follows from $d_N (z) \ne 0$ and the expressions of $b_N, c_N$). Using the fact that
$u_0(n+1, z) u_1(n,z)-u_0(n,z) u_1(n+1,z)=1$ for any $n$ (in particular, for $n=N$), and
\eqref{g9}--\eqref{g10b}, it is easy to calculate $D_N (z)$:
\begin{equation}\label{g104}
D_N(z) \; = \; \frac{d_N (z)}{s_2 (z) u_1(N,z)-u_1 (N+1,z)} \; = \;
\frac{b_N (z)}{s_2 (z) u_0(N,z)-u_0 (N+1,z)}.
\end{equation}
 As observed above, the solutions $u_0, u_1$ are the same for $H, H_N$ if $n \le
N+1$. It follows from \eqref{g8}, where $c_1=0$ and $c_2=D$, that
$$
\langle R_N (z) \delta_1, \delta_n \rangle =
D_N (z) s_2(z)^{n-N}, \ n
\ge N.
$$
The result of the lemma follows now directly from \eqref{g102} and \eqref{g104}.
\hfill $\Box$

\begin{lemma}\label{lem3}
For $z=E+i\e$ with $E \in [-K,K]$ and $0 < \e \le 1$, for $N \ge 3$, we have
$$
M_r(N,z) \le C(K)\e^{-4}\left( \max_{3 \le n \le N} \|\Phi(n,z)\|^2 \right) ^{-1}.
$$
\end{lemma}

\noindent {\bf Proof.}
The second bound of Lemma~\ref{lem1} and the bound \eqref{g6a} of Lemma~\ref{lem2} yield
\begin{align}
M_r(N,z) & \le A(K) \e^{-4} |s_2 (z) u_0(N,z)-u_0(N+1,z)|^{-2}
\sum_{k=0}^\infty |s_2 (z)|^{2k} \label{g12a} \\
& \le B(K) \e^{-4} |s_2 (z) u_0(N,z)-u_0(N+1,z)|^{-2} \label{g12}
\end{align}
with uniform $B(K)$, since $|s_2 (z)|<1$. Then, one has
$$
M_r(N,z) \le C(K)\e^{-4} \left( |u_0(N,z)|^2+|u_0(N+1,z)|^2 \right) ^{-1},
$$
using the fact that $u_0$ is exponentially decreasing in $N$, the product $u_0(N,z)u_0(N+1,z)$ only adds some universal constant to $C(K)$.
Using \eqref{g6b}, we can prove a similar bound with $u_0$ replaced by $u_1$ and therefore obtain
$$
M_r(N,z) \le C(K) \e^{-4} \|\Phi(N,z)\|^{-2}.
$$
Since $M_r(n,z)$ is decreasing in $n$, the asserted bound follows.
\hfill $\Box$

\hspace{2mm}

\noindent {\bf Proof of Theorem~\ref{main}.}
The assertion is an immediate consequence of \eqref{g4} and Lemma~\ref{lem3} (and the
analogous results on the left half-line). As sequences $a$ and $b$ in \eqref{sturmoff} are symmetric the same can be done for $\langle P_l(N,T) \rangle$.  Note that we can replace $[3,N]$ by $[1,N]$
since this modification only changes the $K$-dependent constant.
\hfill $\Box$

\hspace{2mm}

We now show the same result for non time-averaged quantities.
As in \cite{damtche}, the key is to replace the Parseval formula by a Dunford functional calculus (sometime called Riesz-Dunford). 

\begin{lemma}\label{lemmadunford}
For every $n \in \ZM, t \in \RM,$ and positively oriented simple closed contour $\gamma$ in $\CM$ that is such that the spectum of $H$ lies inside $\gamma$, we have
$$
\langle e^{-itH}\delta_1, \delta_n \rangle = -\frac{1}{2\pi i} \int_\gamma e^{-itz} \langle (H-z)^{-1} \delta_1, \delta_n \rangle dz.
$$
.
\end{lemma}

\noindent {\bf Proof}
This is a direct consequence of Dunford functional calculus, see \cite{refdunford1, refdunford2, refdunford3}.
\hfill $\Box$

\begin{lemma}\label{lemma7}
Suppose $H$  defined in \eqref{jacobidef} and $a$ and $b$ in \eqref{sturmoff} and $K \geq 4$ is such that $\sigma(H) \subset [-K+1, K-1]$. Then,
$$
 P(N,t)  \lesssim exp(-cN) + \int_{-K}^{K} \sum_{n \geq N} |\langle (H-E-it^{-1})^{-1} \delta_1,\delta_n \rangle |^2  dE.
$$
\end{lemma}

\noindent {\bf Proof.}
For any $t>0$,  consider the following contour $\gamma = \gamma_1 \cup \gamma_2 \cup \gamma_3 \cup \gamma_4$,
where 
$$
\gamma_1 = \{  E + iy \;{\bf :}\;  E \in [-K,K],\; y \;=\; t^{-1} \},
\quad
\gamma_2 = \{  E + iy \;{\bf :}\; E\;=\;-K,\; y  \in [-1, t^{-1}] \},
$$
$$
\gamma_3 = \{  E + iy \;{\bf :}\; E \in [-K,K],\; y \;=\;-1 \},
\quad
\gamma_4 = \{  E + iy \;{\bf :}\; E\;=\;K,\; y  \in [-1, t^{-1}] \}.
$$
Notice that the spectrum of $H$ lies within this contour and that for $z\in \gamma$,  $\Im  m(z) \leq t^{-1}$ and thus $|e^{-itz} | \leq e$. Lemma \ref{lemmadunford} then implies,
$$
|\langle e^{-itH}\delta_1,\delta_n \rangle | \lesssim \sum_{j=1}^4 \int_{\gamma_j}  |\langle (H-z)^{-1} \delta_1, \delta_n \rangle | |dz|.
$$ 
If $z \in \gamma_2 \cup \gamma_3 \cup \gamma_4$, then we can again apply Combes-Thomas estimates and bound by $C\exp(-dN)$.
The integral over $\gamma_1$ can be estimated using Cauchy-Schwarz inequality:
$$
\hspace{2cm}
\left( \int_{\gamma_1}  |\langle (H-z)^{-1} \delta_1, \delta_n \rangle | |dz| \right)^2 \leq C(K) \int_{-K}^K   |\langle (H-E-\tfrac{i}{t})^{-1} \delta_1, \delta_n \rangle |^2 dE. \hspace{2.5cm} \Box
$$

\begin{theo}\label{main2}
Suppose $H$  defined in \eqref{jacobidef} and $a,b$ in \eqref{sturmoff}, and $K \geq 4$ is such that $\sigma(H) \subset [-K+1, K-1]$. Then the outside probabilities can be bounded from above in terms of transfer matrix norms as follows:
$$
 P(N,t) \lesssim \exp(-dN) +t^4 \int_{-K}^{K} \left( \max_{1\leq n \leq N} \| \Phi (n,E + \tfrac{i}{t}) \|^2 \right)^{-1}dE.
$$
\end{theo}

\noindent {\bf Proof.}
The proof starts by using Lemma \ref{lemmadunford} and \ref{lemma7} instead of Parseval formula. All the steps  are then the same that in the proof of the Theorem \ref{main}.
\hfill $\Box$

\section{Application to off-diagonal Fibonacci dynamic}

We apply Theorem 2 and 3 for Fibonacci off-diagonal model and show a non trivial dynamical upper bound. Namely we show the following Theorem:

\begin{theo}\label{thprincipal}
Consider the off-diagonal Fibonacci Hamiltonian, that is the operator defined in \eqref{jacobidef} and \eqref{sturmoff} with $\beta = \frac{\sqrt{5}-1}{2}$. Assume $c = c^a >8$, then 
$$
\widetilde{\alpha}_u^+ \leq  \frac{2 \log \tfrac{\sqrt{5}-1}{2}}{\log \xi_c},
$$
where  $\xi_c = c-2 + \sqrt{c^2-4c+1}$.

The same holds for non time-averaged dynamical exponent $\alpha^+_u$.
\end{theo}

\begin{rem}
Picking $c$ large enough, one obtains a non trivial bound, that is better that the so-called ballistic bound $1$.
\end{rem}

Considering Fibonacci special case symplifies greatly the trace map evolution, since one has to consider only one sequence of trace as $z_n = x_{n+1}$. Indeed evolution of the tracemap reduces to the following:
$$
x_{n+1} \; = \; x_nx_{n-1} -x_{n-2}
$$
with 
\begin{equation}
\label{condinitfibo}
x_{-1}= \frac{\lambda_1^2+\lambda_2^2}{\lambda_1 \lambda_2}, \quad x_0 = \frac{E}{\lambda_2}, \quad x_1 = \frac{E}{\lambda_1}.
\end{equation} 
Denote by $F_k$ the Fibonacci sequence.
A direct application of Lemma \ref{explosion1} yields in this case to either $|x_0|\leq 2$ or $|x_1|\leq 2$ and implies $N_0(E) \leq 1$ uniformly in $E$. This allow to extend  description in term of periodic band spectrum.
We  recall  some results, classical in on-diagonal case (see Appendix B), and  extended to off-diagonal Fibonacci case since  proof   depend mostly on the trace map application evolution (see \cite{Janine}). For the same technical reasons that in on-diagonal case, we should suppose that $c>4$.

\begin{defini}
Define a band $B_k \subset \sigma_{k,0}$ to be of type A if  $B_k \subset \sigma_{k-1,0}$ and a band of type B if  $B_k \subset \sigma_{k-2,0}$. 
\end{defini}

This definition exhausts all possibilities as seen in the following lemma:

\begin{lemma}\label{bandtype}
Let $c>4$, and $k \geq 2$. Then 
\begin{itemize}
\item[(i)] Each type A band $B_k \subset \sigma_{k,0}$ contains exactly one type B band  $B_{k+2} \subset \sigma_{k+2,0}$ and no other bands from $\sigma_{k+1,0}$ and $\sigma_{k+2,0}$.
\item[(ii)] Each type B band $B_k \subset \sigma_{k,0}$ contains exactly one type A band  $B_{k+1} \subset \sigma_{k+1,0}$ and two type B bands from $\sigma_{k+2,0}$ positionned around $B_{k+1}$  and no other bands from $\sigma_{k+1,0}$ and $\sigma_{k+2,0}$.
\end{itemize}
\end{lemma}

\noindent {\bf Proof.}
The proof inpired from diagonal case \cite{r1} can be found in \cite{Janine}.
\hfill $\Box$

\vspace{0.2cm}
We recall also some estimates on the band sizes given in \cite{Janine}. Those estimates were there used to compute the fractal dimension of the spectrum.

\begin{lemma}\label{bandsize}
Let $c>8$, and $k \geq 2$. Then, with $\xi_c = c-2 + \sqrt{c^2-4c+1}$, we have the following inequalities:

\begin{itemize}
\item[(i)] For any type A band $B_{k+1} \subset \sigma_{k+1,0}, E \in B_{k+1}$ implies 
$$
\xi_c \leq \left|  \frac{x_{k+1}' (E) }{x_k' (E)} \right| \leq
2c+7.
$$ 
\item[(ii)] For any type B band $B_{k+2} \subset \sigma_{k+2,0}, E \in B_{k+2}$ implies 
$$
\xi_c \leq \left|  \frac{x_{k+2}' (E) }{x_k' (E)} \right| \leq
2(2c+7).
$$ 
\end{itemize}
\end{lemma}

By now, we consider the periodic approximants spectrum in $\mathbb{C}$.
$$
\sigma_{k,0}^\delta = \{z \in \mathbb{C} : |x_k(z)| \leq 2 + \delta\}.
$$

All the properties keep true replacing $\sigma_{k,0}$ by $\sigma_{k,0}^\delta$ for some small enough fixed $\delta$ (recall the occurence of $\delta$ in Lemma \ref{lemma5}), in particular statement \eqref{Binfini}. A condition on $c$  should be added to keep Fricke-Vogt invariant, $c > \lambda(\delta) = [12(1+\delta)^2+8(1+\delta)^3+4]^{1/2}$.

The following Proposition states, due to classical Koebe distortion theorem, the height of the set $\sigma_k^\delta$  is almost the same that its length.

\begin{proposi}\label{prop2}
If $k \geq 3$, $\delta >0$ and $c> \max (8, \lambda(\delta))$ then there exist constants
$c_\delta$,$d_\delta>0$ such that
$$
\bigcup_{j=1}^{F_{k-1}} B ( x_k^{(j)},r_k) \subseteq \sigma_{k,0}^\delta
\subseteq \bigcup_{j=1}^{F_{k-1}} B ( x_k^{(j)},R_k)
$$
where $\{x_k^{(j)}\}_{1 \leq j \leq q_{k-1}}$ are the zeros of $x_k$,
$r_k$ and $ R_k$ are the sizes of respectively the smallest and the largest band in $\sigma_k^\delta$.
\end{proposi}

\noindent {\bf Proof.}
The proof is a direct consequence of properties of the functions $x_k(E)$ which are proper and continuous as polynomials in $E$ and the distorsion theorem of Koebe. See \cite{dam1, Laurentarticle, LaurentThesis} for details.
\hfill $\Box$

\vspace{0.5cm}

We have now all the  required tools to finish the proof of the theorem \ref{thprincipal}.

\vspace{0.5cm}

\noindent {\bf Proof of Theorem \ref{thprincipal}.}
As $x_k^{(j)}$ are real, we have with Proposition \ref{prop2} 
$$
\sigma_{k}^\delta \subseteq \{ z \in \CM \text{:} |\Im m\;  z | < R_k\} \subseteq \{z \in \CM \text{:} |\Im m \;
 z | <d
F_k^{-\gamma(c)}\}.
$$
for a suitable $\gamma (c)$ .
This implies 
\begin{equation}
\label{eq_expl}
\sigma_{k,0}^\delta \cup \sigma_{k+1,0}^\delta \subseteq \{z \in \CM \text{:} |\Im m\; z | < d
F_k^{-\gamma(c)}\}.
\end{equation}

Let us precise how to choose  $\gamma(c)$.

From Lemma \ref{bandtype} and Lemma \ref{bandsize}, it is easy to bound $R_k$:
$$
R_k \leq \xi_c^{-k/2}
$$

We should have
$
R_k < d F_k^{-\gamma(c)}
$
so a suitable $\gamma$ can be chosen  by taking:
$$
\gamma (c) \leq \limsup_k  \frac{k\log \xi_c}{2  \log F_k }.
$$
Remarking that $F_k$ behave like $\left( \frac{\sqrt{5}-1}{2}\right)^k$ leads to
$\gamma(c) \leq  \frac{\log \xi_c}{2\log \tfrac{\sqrt{5}-1}{2}} $.

For $\varepsilon = \Im m\; z >0$, we get a lower bound for $|x_n(E+i\varepsilon)|$ uniform in  $E \in [-K,K] \subset \RM$. For a fixed
$\varepsilon>0$, we choose $k$ such that $ d
F_k^{-\gamma(V)} < \varepsilon$. With (\ref{eq_expl}), this shows $|x_k (E +
i\varepsilon)| > 2+\delta$ and $|x_{k+1} (E +i\varepsilon)| > 2+\delta$. As $|x_{1}(E +
i\varepsilon)|\leq 2+\delta$ or $|x_{0}(E +
i\varepsilon)|\leq 2+\delta$ from Lemma \ref{explosion1}, we can apply Lemma \ref{lemma5}
and thus
$$
|x_j| \geq e^{\log(1+\delta) F_{j-k}} +1 \hspace{5mm} \forall j > k.
$$

All this motivates the following definitions:
\begin{defini}
 For $\delta >0, T>1$, denote by $k(T)$  the unique
integer with
$$
\frac{F_{k(T)-1}^{\gamma(c)}}{d} \leq T \leq \frac{F_{k(T)}^{\gamma(c)}}{d}
$$
and let
$$
N(T) = F_{k(T) + \lfloor \sqrt{k(T)} \rfloor}.
$$
\end{defini}

For every  $\nu > 0$, there is a constant $C_\nu >0 $ such that
\begin{equation}{\label{estiN(T)}}
N(T) \leq C_\nu T^\frac{1}{\gamma (V)} T^\nu.
\end{equation}
Applying Theorem \ref{main} and above estimate, we get
\begin{align*}
P_r (N(T),T) &\lesssim \exp(-cN(T)) + T^3 \int_{-K}^K \left( \max_{1\leq F_n \leq N(T)}
\left\| M_n( E + \tfrac{i}{T})
\right\|^2 \right)^{-1} dE), \\
	     &\lesssim \exp(-cN(T)) + T^3 e^{-2\log(1+\delta) F_{\lfloor
\sqrt{k(T)} \rfloor}}. 
\end{align*}

From this bound, it is clear that  $P_r (N(T),T)$ goes to zero faster than any inverse power of $T$ and thus
$$
\widetilde{\alpha}_u^+ = \frac{1}{\gamma(c)} + \nu
$$
with  $\nu$ arbitrary small.

One complete the proof without time averaging, using Theorem \ref{main2} instead of Theorem \ref{main}.
\hfill $\Box$

\section{Fractal dimension of the on-diagonal model spectrum }

We now consider $H$ defined with \eqref{jacobidef} and \eqref{sturmon}.
Denotes by $N(\varepsilon)$ the minimal number of  balls of diameter at most $\varepsilon$ one need to cover $\sigma$, then the upper (and lower) box-counting dimension are defined respectively by
$$
\dim_B^+(\sigma) = \limsup_{\varepsilon \to 0}\frac{\ln N(\varepsilon)}{\ln \varepsilon},
\quad
\dim_B^-(\sigma) = \liminf_{\varepsilon \to 0}\frac{\ln N(\varepsilon)}{\ln \varepsilon}.
$$
When the limit exists, one denotes it simply by $\dim_B(\sigma)$.

We give a lower bound of minimal number of balls of some explicit decreasing scale $\varepsilon_k$, needed to cover the spectrum.  The first idea to cover the spectrum can be  to take into account all the bands and take  as a scale the smallest band size, but this is a bad idea because this minimal length  decreases faster than the number of bands grows. A better idea will be to count how many bands have a length of order  $\varepsilon_k \approx\lambda_1^{-k}$.
 This yields to a better lower bound for the box-counting dimension of the spectrum associated to a large class of irrational number.

\begin{theo}\label{thbornesur}
Set  $C_k= \frac{3}{k} \sum_{j=1}^k \log (a_j +2)$. We have for
any irrational number  $\beta$ verifying
$C = \limsup C_k <+\infty$ and $\lambda_1>20$:
\begin{equation}\label{bornesur}
\dim_B^+ (\sigma) \geq \frac{\ln \left( \frac{\sqrt{5}+1}{2}\right)}{C + \ln (\lambda_1+5)}
\end{equation}
where $\sigma$ is the spectrum of $H$.
\end{theo}

\begin{rem}
 The diophantine condition $C < +\infty$ is Lebesgue measure $1$ (see Appendix A). In particular, it is true for random numbers and degree 2 algebraic numbers. 
\end{rem}

As in off-diagonal Fibonacci, one can label bands (see appendix B for precise definitions).
The following Lemma give precise statement of the counting idea.
\begin{lemma}\label{lemrec}
Denote by $n_{k,I}, n_{k,II}$ and  $n_{k,III}$ the number  of bands of type  I, II and  III in \break $\sigma_{k,1}, \; \sigma_{k+1,0}, \; \sigma_{k+1,0}$  with a length  greater than 
$\varepsilon_k =4 \Pi_{j=1}^k (\lambda_1+5)^{-1} (a_j +2)^{-3}$.

For all  $k$, we have the following induction:
\begin{equation}\label{evol1}
n_{k+1,I} \;=\; (a_{k+1} +1)n_{k,II} + a_{k+1} n_{k,III},
\end{equation}
\begin{equation}\label{evol2}
n_{k+1,II} \;\geq\; \mathds{1}_{\{a_{k+1} \leq 2\}} n_{k,I},
\end{equation}
\begin{equation}\label{evol3}
n_{k+1,III}\; =\; a_{k+1} n_{k,II} + (a_{k+1}-1) n_{k,III},
\end{equation}
with initial conditions  $n_{0,I}=1,\quad n_{0,II}=0,\quad  n_{0,III}=1$.

Moreover, 
$$
n_{k,II} \neq 0 \bigvee  n_{k,III} \neq 0,
\quad
n_{k,I} \neq 0,
\quad
n_{k,I} > n_{k,III},
$$
and
\begin{equation}\label{ngrowth}
n_{k,II} + n_{k,III} > \left(\frac{\sqrt{5}+1}{2}\right)^{k}.
\end{equation}

\end{lemma}

\noindent {\bf Proof.}
The induction relation is obvious with Lemma \ref{raymond} and Theorem \ref{thchinois} from Appendix B.

The two first properties are made by induction.
Initial conditions give level $0$.
 Assume it is true at level $n$, then  as $a_{k+1} > 0$,  $n_{k,II} \neq 0 \bigvee  n_{k,III} \neq
0$, implies   $n_{k+1,I} \neq 0$.
For the second part, if  $a_{k+1} \leq 2$ then $n_{k+1,II} \neq 0$, else $a_{k+1}
> 2$ implies $n_{k+1,III} \neq 0$.

To prove
$
n_{k,I} > n_{k,III},
$
it suffices to see that
$
n_{k,I} \geq  n_{k,III} + n_{k-1,II} + n_{k-1,III}.
$

Denote by $n_k$ the sum of $n_{k,II}$ and $n_{k,III}$.
For the last property, we argue that for $k>1$,
\begin{equation}\label{eqsuite}
n_{k}  \geq 2 n_{k-2} + n_{k-3} .
\end{equation}
It is easy to verify from \eqref{eqsuite} that the sequence $n_k$ growth verifies \eqref{ngrowth}.

We now show \eqref{eqsuite}.
Using  \eqref{evol1}-\eqref{evol3}, we get
$$
n_{k,II} = [(a_{k-1} +1) n_{k-2,II} + a_{k-1}n_{k-2,III}] \mathds{1}_{\{a_k \leq 2\}},
$$
$$
n_{k,III} = (a_k-1) (a_{k-1} n_{k-2,II} + (a_{k-1}-1)n_{k-2,III}) + a_k
n_{k-2,I} \mathds{1}_{\{a_{k-1} \leq 2\}}.
$$

We distinguish all the cases depending on the values of $a_k$ and
$a_{k-1}$ in the following table.
\\
\\
\begin{tabular}{|c|c|c|c|}
\hline
$n_k \geq$  & $a_{k-1}=1$ & $a_{k-1}=2$ & $a_{k-1}=\gamma>2$ \\
\hline
$a_{k}=1$ & $2n_{k-2} + n_{k-3}$  & $3n_{k-2} + n_{k-3}$  & $ \gamma n_{k-2}$ \\
\hline
$a_{k}=2$  & $3n_{k-2} +2 n_{k-3}$  & $5n_{k-2} +2 n_{k-3}$  & $(2\gamma-1) n_{k-2}$\\
\hline
$a_{k}=\lambda>2$  & $(\lambda-1)n_{k-2} + \lambda n_{k-3}$  & $2(\lambda-1)n_{k-2} + \lambda n_{k-3}$  & $(\lambda-1)(\gamma-1)n_{k-2} + (\lambda-1) n_{k-3}$\\
\hline
\end{tabular}

\vspace{0.5cm}
Therefore, the slowest growth for the quantity $n_{k,II} + n_{k,III}$ is for $a_k = a_{k-1} =1$ which ends the proof. 
\hfill $\Box$

\hspace{5mm}

\noindent { \bf Proof of theorem \ref{thbornesur}}
With  Lemma \ref{lemrec}, we obtain a bound for $n_{k,II} +
n_{k,III}$, that is the number of bands of length at least $\varepsilon_k$. 
Then by definition of box-counting dimension, we have
$$
\dim_B^+ (\sigma) \geq \liminf_k \frac{\ln (n_{k,II} +
n_{k,III})}{-\ln \varepsilon_k}
$$
and the stated result.
\hfill $\Box$

\section{Dimension almost sure}

In this part, we consider only random numbers, that is number with $C = 5.04\dots$ (see Corrollary 1 in Appendix A). This class of course is included in the class of the previous section and is still Lebesgue measure $1$.
This first Lemma is purely technical and  link the ergodicity of a random continued fraction expansion (see Appendix A) with our two step setting.

\begin{lemma}\label{lempropor}
Denote $E(\lambda,\gamma)$ the event $\{(a_{2k},a_{2k+1}) = (\lambda,\gamma)\}$ for some $k>0$. For every natural $\lambda,\gamma$, we have  $P(E(\lambda,\gamma))=\frac{\ln\left( 1 + \frac{1}{\lambda(\lambda+2)}\right)\ln\left( 1 + \frac{1}{\gamma(\gamma+2)}\right)}{(\ln(2))^2}$.
For all $\varepsilon>0$, there exists $N_0$ big enough, such that for every $N>N_0$, we have the probability:
$$
P\left( \text{At least\hspace{1mm}} \left(P(E(\lambda,\gamma))-\varepsilon \right)N \hspace{1mm}\text{couples verify \hspace{1mm}} E(\lambda,\gamma)\right) =1.
$$
\end{lemma}

\noindent { \bf Proof.}
Denote $A(r)$ the event 
$$
A(r)=\{\text{Among the \hspace{1mm}} n \hspace{1mm} \text{ first} \hspace{1mm}  \text{there is exactly} \hspace{1mm} r \hspace{1mm} \text{couples verifying }  E(\lambda,\gamma)\}.
$$

\noindent Since  $a_{2k}$ and $a_{2k+1}$ are independant random variables, we have $P(E(\lambda,\gamma))=\frac{\ln\left( 1 + \frac{1}{\lambda(\lambda+2)}\right)\ln\left( 1 + \frac{1}{\gamma(\gamma+2)}\right)}{(\ln(2))^2}$ and set for convienience $\frac{1}{B} = P(E(\lambda,\gamma))$, we have 
$$
P(A(r)) = C_n^r \frac{1}{B^r} \left(1-\frac{1}{B}\right)^{n-r}.
$$

Then the event $\{ \text{There is  less than \hspace{1mm}} r \hspace{1mm}\text{couples verifying \hspace{1mm}} E(\lambda,\gamma)\}$ is $\bigcup_{k=0}^r A(k)$ and

\begin{align}
P\left(\bigcup_{k=0}^r A(k)\right) &= \sum_{k=0}^r  C_n^k \frac{1}{B^k} \left(1-\frac{1}{B}\right)^{n-k} \notag \\
                     \label{eqP}    &\leq C_n^r \left(\frac{B-1}{B}\right)^n \left(\frac{r+1}{n-r}\right)^r \sum_{k=0}^r \left( \frac{n-r}{(B-1)(r+1)}\right)^k 
\end{align}
Using $C_n^{k+1}= \frac{n-k}{k+1} C_n^k \geq \frac{n-r}{r+1} C_n^k$ o compare $C_n^k$ and $C_n^r$.  
Then assuming that $r<\frac{n}{B}$, one shows that 
$$
\frac{n-r}{(B-1)(r+1)} \leq \frac{n-n/B}{(B-1)(n/B+1)}=\frac{n}{n+B} <1.
$$

It follows that the remaining sum in (\ref{eqP}) is geometric and is bounded by some constant $K$.
 
Using then Stirling formula, one can estimate $C_n^r$:
$$
C_n^r \sim \frac{n^n}{(n-r)^{n-r}r^r}.
$$

This yields to the following
$$
P\left(\bigcup_{k=0}^r A(k)\right) \leq K \left( \frac{n}{n-r} \right)^n\left( \frac{B-1}{B} \right)^n \left( \frac{r+1}{r} \right)^r.
$$
If $r \leq \left(\frac{1}{B}-\varepsilon\right)n$, for some $\varepsilon>0$, then  the right term in the above expression tends to $0$, which ends the proof.
\hfill $\Box$

\begin{theo}\label{bornepresquesure}
For almost all irrational number $\beta$, an arbitrary small $\nu >0$ and $\lambda_1>20$:
$$
\dim_B^+ (\sigma)  \geq \frac{D-\nu}{C+\ln (\lambda_1 +5)}
$$
with $D=1.0382\dots$ 
\end{theo}

\begin{rem}
$D$ is obtained as the limit of an explicit converging series. Value of $D$ is roughly twice bigger than the bound \eqref{bornesur} in Theorem \ref{thbornesur}. It is also worthy to be compare it with optimal bound for golden and silver mean $D = 0.88\dots$ {\rm \cite{tche2,chin2}}.  
\end{rem}

\noindent {\bf Proof.}
First, simplify the recursion relation in Lemma \ref{lemrec} by  ignoring the smallest term, and thus obtain:
$$
n_{2k,II}+ n_{2k,III} \geq c_{2k} \left( n_{2k-2,II}+ n_{2k-2,III} \right)
$$
where $c_{2k}$ depend only on the values of the couple $(a_{2k},a_{2k-1})$ according to the table:
\begin{center}
\begin{tabular}{|c|c|c|c|}
\hline
$c_{2k}(a_{2k},a_{2k-1})$ & $a_{2k} =1$ & $a_{2k}=2$ & $a_{2k}=\lambda>2$ \\
\hline
$a_{2k-1}=1$ & 2 & 3 & $\lambda -1$\\
\hline
$a_{2k-1}=2$ & 3 & 5 & $2(\lambda-1)$ \\
\hline
$a_{2k-1}=\gamma >2$ &$\gamma$& $2\gamma-1$ & $(\lambda-1)(\gamma-1)$\\
\hline
\end{tabular}
\end{center}

Using now Lemma \ref{lempropor} and probability value for $a_k$ recall in Corollary \ref{thkhin}  in Appendix A,  one has for all $k>0$ and almost all $\beta$:
\begin{multline*}
n_{2k,II}+ n_{2k,III} \geq 
2^{k(P(E(1,1))-\varepsilon)} 3^{k(P(E(2,1))+P(E(1,2))-\varepsilon)}  5^{k(P(E(2,2))-\varepsilon)} \prod_{\lambda=3}^n \lambda^{k(P(E(1,\lambda)-\varepsilon)}  \times \\
\prod_{\lambda=3}^n (\lambda-1)^{k(P(E(\lambda,1)-\varepsilon)}  
\prod_{\lambda=3}^n 
(2\lambda-1)^{k(P(E(2,\lambda)-\varepsilon)} 
\prod_{\lambda=3}^n 
(2\lambda-2)^{k(P(E(\lambda,2)-\varepsilon)} \times \\
\prod_{\lambda,\beta=3}^n ((\lambda-1)(\beta-1))^{k(P(E(\beta,\lambda)-\varepsilon)}.\hspace{15mm} 
\end{multline*}

In the previous equation, we replace all the $\varepsilon$ constants by only one constant denoted also $\varepsilon$ and we also take the same $n$ in all the products. We can do those simplifications since we consider only a finite number of terms.
Taking the logarithm on the above expression, one obtains:
\begin{multline}\label{eq18}
\frac{\ln (n_{2k,II}+ n_{2k,III})}{2k(1-\varepsilon)} \geq \frac{1}{2 (\ln 2)^2}
\left(  \ln (2)\left(\ln \frac{4}{3}\right)^2 + 2\ln (3)\left(\ln \frac{9}{8}\right)\left(\ln \frac{4}{3}\right) + \right. \\
 \ln (5)\left(\ln \frac{9}{8}\right)^2 + 
\sum_{\lambda =3}^n \big(\ln(\lambda)+\ln(\lambda-1)\big)\left(\ln \frac{4}{3}\right) \ln\left(1+\frac{1}{\lambda(\lambda+2)}\right)+  \\
\sum_{\lambda =3}^n \big(\ln(2\lambda-1)+\ln(2\lambda-2)\big)\left(\ln \frac{9}{8}\right) \ln\left(1+\frac{1}{\lambda(\lambda+2)}\right)+  \\
\left. \sum_{\lambda,\beta=3}^n \big( \ln (\lambda-1) + \ln(\beta-1) \big)
\ln\left(1+\frac{1}{\lambda(\lambda+2)}\right)
\ln\left(1+\frac{1}{\beta(\beta+2)}\right)\right) .
\end{multline}

We can rewrite the last double sum:
\begin{align*}
\sum_{\lambda=3}^n \sum_{\beta =3}^n &\big( \ln (\lambda-1) + \ln(\beta-1) \big)
\ln\left(1+\frac{1}{\lambda(\lambda+2)}\right)
\ln\left(1+\frac{1}{\beta(\beta+2)}\right) \\ 
&= 2 \sum_{\lambda=3}^n \sum_{\beta =3}^n \ln (\lambda-1) \ln\left(1+\frac{1}{\lambda(\lambda+2)}\right)
\ln\left(1+\frac{1}{\beta(\beta+2)}\right) \\
&= 2 \sum_{\lambda=3}^n \ln (\lambda-1) \ln\left(1+\frac{1}{\lambda(\lambda+2)}\right)
 \sum_{\beta =3}^n \ln\left(1+\frac{1}{\beta(\beta+2)}\right) \\
&= 2 \sum_{\lambda=3}^n \ln (\lambda-1) \ln\left(1+\frac{1}{\lambda(\lambda+2)}\right)\left( \ln \frac{4}{3} \right).
\end{align*}

As all the series in $n$ have positive terms and are convergent, we can bound from below by taking the limit minus some $\nu>0$ arbitrary small.

The limit of r.h.s  of (\ref{eq18}) can be numerically compute and gives $D= 1.0382\dots$.
\hfill $\Box$

\section*{Appendix A: A theorem by Khintchin}

We recall here some of the ergodic properties of the continued fraction expansion process.

\begin{theo}{\rm \cite{khinchin}}\label{thkhint}
Suppose that $f(r)$ is a non-negative  function of a  natural argument $r$,  (r=1,2,\dots), and suppose that there exist positive contants $C$ and $\delta$ such that 
$$
f(r) < Cr^{\frac{1}{2}-\delta}.
$$
Then, for all numbers in the interval $(0,1)$, with the exception of a set of measure zero,
$$
\lim_{n \to \infty} \frac{1}{n} \sum_{k=1}^n f(a_k) = \sum_{r=1}^\infty f(r)\frac{\ln\left( 1+ \frac{1}{r(r+2)}\right)}{\ln 2}.
$$
\end{theo}

\begin{coro}\label{thkhin}
For almost all $\beta$ with respect to Lebesgue measure,
$$
C= \limsup_k  \frac{3}{k}\sum_{j=1}^k \log (a_j+2) =5.04\dots
$$

Moreover, the density of the number $i$ in the sequence $\{a_k\}_k$ is almost surely
$$
d(i) = \frac{\ln\left( 1 + \frac{1}{r(r+2)}\right)}{\ln 2}.
$$
\end{coro}

\noindent {\bf Proof.}
Apply Theorem \ref{thkhint} with $f(r) = \ln (r+2)$ and $f(r) = \mathds{1}_i(r)$ respectively.
\hfill $\Box$

\section*{Appendix B: Gap labelling and band estimates}

We recall precise information on periodic on-diagonal spectrum sets. 

\begin{defini} For a given $k$, we call

type I gap : a band of $ \sigma_{k,1}$ included in a band of $ \sigma_{k,0}$ and therefore in a gap of $ \sigma_{k+1,0}$,

type II band : a band of $ \sigma_{k+1,0}$ included in a band of $ \sigma_{k,-1}$ and in a gap of $ \sigma_{k,0},$

type III band : a band of $ \sigma_{k+1,0}$ included in a band of $ \sigma_{k,0}$ and in a gap of $ \sigma_{k,1}$.
\end{defini}

These definitions exhaust all the possible configurations:

\begin{lemma} {\rm \cite{r1}} \label{raymond}
At a given level $k$,

\noindent {\rm (i)} a type I gap contains an unique type  II band of $\sigma_{k+2,0}$.

\noindent {\rm (ii)} a type II band contains $(a_{k+1}+1)$ bands of type I of $\sigma_{k+1,1}$. They are alternated with $(a_{k+1})$ type III bands of $\sigma_{k+2,0}$

\noindent {\rm (iii)} a type III band contains $(a_{k+1})$ bands of type I of $\sigma_{k+1,1}$. They are alternated with $(a_{k+1}-1)$ type III bands of $\sigma_{k+2,0}$
\end{lemma}

Let  $\mathcal{A} = \{I,II,III\}$ be a three letters alphabet. For each band $B$ of spectrum at level $k$, correspond an unique word $i_0i_1\dots i_k \in \mathcal{A}^{n+1}$ such that $B$ is  a band of type $i_k$ included in a band of type
$i_{k-1}$ at level $k-1$,...,included in a band of type $i_0$ at level 0.
This word will be called the $index$ of $B$. More than one band can have the same index.
Let $T_n = (t_{i,j}(n))_{3*3}$ be a sequence of matrix and $\tau = i_0i_1\dots i_k$ an index, we define:
$$
L_\tau (T) = t_{i_0,i_1}(1)t_{i_1,i_2}(2)\dots t_{i_{k-1},i_k}(k).
$$

\begin{theo} {\rm \cite{chin1}} \label{thchinois}
 If $\beta = [a_1,a_2...]$ is an irrational number in  $(0,1)$ and $H$ defined 
as in \eqref{jacobidef} and \eqref{sturmon} with $\lambda_1> 20$ then any band   $B$ of index $\tau$ verifies,
$$
4 L_\tau (Q) \leq |B| \leq 4 L_\tau (P)
$$

 where $P=(P_n)_{n>0}$
$$
P_n = \begin{pmatrix}
0 & c_1^{a_n-1} & 0 \\ c_1/a_n & 0 & c_1/a_n \\c_1/a_n & 0 & c_1/a_n
\end{pmatrix}
$$
with $c_1= \frac{3}{\lambda_1-8}$
and $Q=(Q_n)_{n>0}$
$$
Q_n =\begin{pmatrix}
0 & c_2^{a_n -1} & 0 \\ c_2 (a_n +2)^{-3} & 0 & c_2 (a_n +2)^{-3} \\ c_2 (a_n
+2)^{-3} &0 & c_2 (a_n +2)^{-3} \\
\end{pmatrix}
$$
with $c_2= \frac{1}{\lambda_1+5}$.
\end{theo}

\end{document}